\documentclass[Physsubmission, Phys]{SciPost}

\hypersetup{colorlinks,
linkcolor={red!50!black},
citecolor={blue!50!black},
urlcolor={blue!80!black}}

\usepackage[bitstream-charter]{mathdesign}
\def\a{\alpha}

\def\n{\nu}

\def\p{\pi}

\def\r{\rho}

\def\t{\tau}

\def\P{\Pi}

\DeclareSymbolFont{usualmathcal}{OMS}{cmsy}{m}{n}
\DeclareSymbolFontAlphabet{\mathcal}{usualmathcal}

\begin{document}

\begin{center}{\Large \textbf{ \boldmath
Strong coupling at the $\tau$-mass scale from an improved vector isovector spectral function\\}}\end{center}

\begin{center}
Diogo Boito,\textsuperscript{1,2$\star$}
Maarten Golterman,\textsuperscript{3,4}
Kim Maltman,\textsuperscript{5,6}
Santiago Peris,\textsuperscript{4}
\\ Marcus V. Rodrigues,\textsuperscript{1}~and
Wilder Schaaf.\textsuperscript{3,7}
\end{center}

\begin{center}
{\bf 1} Instituto de F\'isica de S\~ao Carlos, Universidade de S\~ao Paulo, CP 369, 13560-970, \\ S\~ao Carlos, SP, Brazil
\\
{\bf 2} University of Vienna, Faculty of Physics, Boltzmanngasse 5, A-1090 Wien, Austria
\\
{\bf 3} Department of Physics and Astronomy, San Francisco State University, \\ San Francisco, CA 94132, USA
\\
{\bf 4} Department of Physics and IFAE-BIST, Universitat Aut\`onoma de Barcelona\\
E-08193 Bellaterra, Barcelona, Spain
\\
{\bf 5} Department of Mathematics and Statistics, York University, Toronto, ON Canada M3J~1P3
\\
{\bf 6} CSSM, University of Adelaide, Adelaide, SA~5005 Australia
\\
{\bf 7} Department of Physics, University of Washington, Seattle, WA 98195
\\
* boito@ifsc.usp.br
\end{center}

\begin{center}
\today
\end{center}

\definecolor{palegray}{gray}{0.95}
\begin{center}
\colorbox{palegray}{\begin{minipage}{0.95\textwidth}
\begin{center}
{\it 16th International Workshop on Tau Lepton Physics (TAU2021),}\\
{\it September 27 – October 1, 2021} \\
\end{center}
\end{minipage}}
\end{center}
\vspace{-1cm}
\section*{Abstract}
{\bf \boldmath
We perform a precise extraction of the QCD coupling at the $\tau$-mass scale, $\alpha_s(m_\tau)$, from a new vector isovector spectral function
which combines ALEPH and OPAL distributions for the dominant channels, $\tau\to\pi \pi^0\nu_\tau$, $\tau\to 3\pi \pi^0\nu_\tau$ and
$\tau\to \pi 3\pi^0\nu_\tau$, with estimates of sub-leading contributions obtained from
electroproduction cross-sections
using CVC, as well as BaBar results for $\tau \to K^-K^0\nu_\tau$. The fully inclusive spectral function thus obtained is entirely based on experimental data, without Monte Carlo input.
From this new data set, we obtain $\alpha_s(m_\tau)=0.3077\pm0.0075$, which corresponds to $\alpha_s(m_Z)=0.1171\pm0.0010$. This analysis can be improved on the experimental side with new measurements of the dominant $\pi\pi^0$, $\pi 3\pi^0$, and $3\pi \pi^0$ $\tau$ decay modes.}

\vspace{1pt}
\noindent\rule{\textwidth}{1pt}
\tableofcontents\thispagestyle{fancy}
\noindent\rule{\textwidth}{1pt}
\vspace{1pt}

\section{Introduction}
\label{sec:intro}

The extraction of the strong coupling, $\alpha_s$, from weighted integrals of inclusive hadronic tau decay data ranks among the most precise determinations of this coupling based on experiment. Since 2008, the $\alpha_s^4$ term in the perturbative QCD description of these inclusive quantities is exactly known~\cite{Baikov:2008jh,Herzog:2017dtz}, which has renewed the interest in this process. Here, we report on the work of Ref.~\cite{Boito:2020xli} where an updated, more precise, extraction of $\alpha_s$ has been performed, using a new vector isovector spectral function.

In order to perform reliable, self-consistent, QCD analyzes of tau decay data, the hadronic $\tau$ decay width is not sufficient --- it is crucial to have information about the spectral functions. Any realistic analysis includes several integrated spectral function moments, which allows for the extraction of $\alpha_s$ and non-perturbative contributions in a combined fit. Obtaining the necessary experimental data is intrinsically difficult since, in principle, the theoretical description in QCD requires fully inclusive spectral functions, with measurements of all subleading hadronic decay channels. The most complete spectral functions have been produced and made publicly available by the ALEPH~\cite{ALEPH:1998rgl,ALEPH:2005qgp} and OPAL~\cite{OPAL:1998rrm} collaborations.\footnote{The spectral functions measured by CLEO~\cite{CLEO:1995nlc} were never made public, to the best of our knowledge.}
Later, a number of updates of the ALEPH spectral functions have been produced, reflecting new experimental information on the different branching ratios and improving the separation in the vector (V) and axial-vector (A) components, among other changes. The most recent version was published in 2013~\cite{Davier:2013sfa}, correcting the correlation matrices that were underestimated in an earlier iteration~\cite{Davier:2008sk}. The OPAL data, in turn, were updated to reflect more recent branching ratio measurements in Ref.~\cite{Boito:2012cr}.

Extractions of $\alpha_s$ from the ALEPH data have smaller experimental uncertainties~\cite{Boito:2014sta}, but extractions from the OPAL data are consistent within errors~\cite{Boito:2012cr}. For the lack of a combined analysis, a weighted average between values obtained from the two data sets, assumed to be uncorrelated, has been quoted as the final result for $\alpha_s$ in Ref.~\cite{Boito:2014sta}. This procedure, however, is not ideal since it does not combine the experimental information used to obtain the integrated moments in an optimal way. As is well established in the dispersive studies of the hadronic vacuum polarization contribution to the muon $g-2$, for example, one should first combine the experimental results following a rigorous statistical procedure and then employ this new combined data set to perform discretized integrals~\cite{Keshavarzi:2019abf,Davier:2019can}. This procedure also quantitatively tests the compatibility of the data sets and reveals potential local discrepancies, should they exist. This is one of the questions addressed in the work of Ref.\cite{Boito:2020xli}.

Another reason for revisiting the $\alpha_s$ extraction from $\tau$ decay data is related to the subleading hadronic channels, to which we often refer as ``residual modes". The ALEPH and OPAL spectral functions include detailed measurements of the decay spectrum of dominant channels. These are the $\pi\pi^0$, $\pi 2\pi^0$, $3\pi$, $\pi3\pi^0$, and $3\pi \pi^0$ tau decay channels. However, these are not sufficient to obtain
fully inclusive spectral functions and other subleading channels, such as $\tau \to \omega(\to {\rm non} \, 3 \pi) \pi^- \nu_\tau$, $\tau \to 6\pi \nu_\tau$, and $\tau \to K^-K^0\nu_\tau$, were included by ALEPH and OPAL using Monte Carlo simulations~\cite{OPAL:1998rrm,ALEPH:2005qgp}. These residual modes, although subleading, do play an important role in QCD studies: removing them makes the spectral function not inclusive enough which can lead to non-physical results. Recently, however, many new experimental results have appeared which allows for a significant improvement in the description of the residual modes in the vector channel.
First, BaBar has measured the spectrum of $\tau\to K^-K^0\nu_\tau$~\cite{BaBar:2018qry}. Second, thanks to a wealth of new experimental measurements of many exclusive-channel cross sections in $e^+e^- \to {\rm hadrons}$~\cite{BaBar:2007ceh,BaBar:2017zmc,Achasov:2016zvn,SND:2014rfi,Achasov:2017kqm,BaBar:2018erh,BaBar:2018rkc,Gribanov:2019qgw,BaBar:2006vzy,CMD-3:2013nph,Achasov:2019nws,CMD-3:2017tgb,Achasov:2019duv,BaBar:2007qju,Achasov:2016eyg} by the CMD-3, BaBar, and SND collaborations, conserved vector current (CVC) relations allow essentially all remaining
vector isovector mode contributions to be obtained using exclusive mode
electroproduction cross-section data (with negligible isospin-breaking corrections, as we discuss below). In Ref.\cite{Boito:2020xli}, this recent experimental information was used to build a new vector isovector spectral function solely based on data, without the use of Monte Carlo simulations for any of the residual modes.

This new vector isovector spectral function has two main features. First, it relies on a combination of ALEPH and OPAL measurements of the dominant vector tau decay channels, namely $\t\to\p^-\p^0\n_\t$, $\t\to2\p^-\p^+\p^0\n_\t$ and
$\t\to\p^-3\p^0\n_\t$, which had never been done in this context. Second, as explained above, all subleading modes are obtained from experiment. This new spectral function has smaller errors and is more inclusive than the ALEPH and OPAL results, since it contains more residual modes, covering 99.95\% of the total vector branching ratio (BF). With this new data set, we have performed an updated analysis of $\alpha_s$ at the $\tau$-mass scale, following the framework of Refs.~\cite{Boito:2011qt,Boito:2012cr,Boito:2014sta}.
Below we discuss in more detail the work of Ref.~\cite{Boito:2020xli} and summarize its main results.

\section{Data combination and the new spectral function}

The first ingredient in the new vector isovector spectral function is the combined ALEPH and OPAL data for the dominant $\tau$ decay channels: $\t\to\p^-\p^0\n_\t$, $\t\to 2\p^-\p^+\p^0\n_\t$, and
$\t\to\p^-3\p^0\n_\t$. Ideally, one would like to combine unit normalized spectral distributions for each channel individually, multiply them by up-to-date values of the respective BFs, and subsequently add them up to build the dominant channel contributions. In practice, however, this procedure is not possible because of the presence of 100\% correlations in parts of the covariance matrices of the $3\pi\pi^0$ and $\pi 3\pi^0$ spectral distributions, which prevents the corresponding covariance matrices from being inverted (a crucial step in the data combination, as we will describe below). Without additional information on how these covariances are obtained and what they represent in terms of statistical and systematic errors, the solution to this problem is to first add all dominant channels, already normalized by the respective BFs, for each experiment separately and later combine the sums of the three dominant channels in a single data set. Since in the sum of the dominant channels the well-behaved $\pi\pi^0$ mode gives the largest contribution, this procedure leads to a well-defined covariance matrix and we therefore employ this route in our data combination.

For the data combination algorithm, we follow closely Refs.~\cite{Keshavarzi:2018mgv,Keshavarzi:2019abf}. In summary, we first define $N_{\rm cl}$ {\it clusters} which roughly play the role of bins for the combined data. Each cluster contains a number of data points. The representative value of $s$ for each cluster, where $s$ is the invariant mass of the final-state hadrons, is given by the weighted average of the $s$ values of the data points contained in that cluster (the weights in this average are the inverses of the data-point uncertainty squared). The description of the combined data set is done by a piece-wise linear function given by the linear interpolation between the values of the spectral function in each cluster, that we denote $\rho^{(m)}$. Extrapolation is used for the few data points whose energy is smaller (larger) than the smallest (largest) cluster energy.

The $N_{\rm cl}$ values of the spectral function at each cluster, $\rho^{(m)}$, are then obtained by minimizing the following $\chi^2$ function
\begin{equation}
\chi^2 = \sum_{i,j=1}^{N_{\rm data}} [\,d_i-R(s_i;\rho)\,]\, (C^{-1})_{ij}\, [d_j-R(s_j;\rho)],
\end{equation}
where $N_{\rm data}$ is the total number of experimental data points, $d_i$ is $i$-th experimental value of the spectral function, and $R(s_i;\rho)$, which plays the role of ``theory" in the minimization, is the value of the combined spectral function at the energy $s_i$, obtained as the linear interpolation between $\rho^{(m)}$ at each cluster center (and using extrapolation at the extremes). Finally, $C$ is the total covariance matrix of the $N_{\rm data}$ experimental data points.

\begin{figure}[t]
\centering
\includegraphics[width=0.45\textwidth]{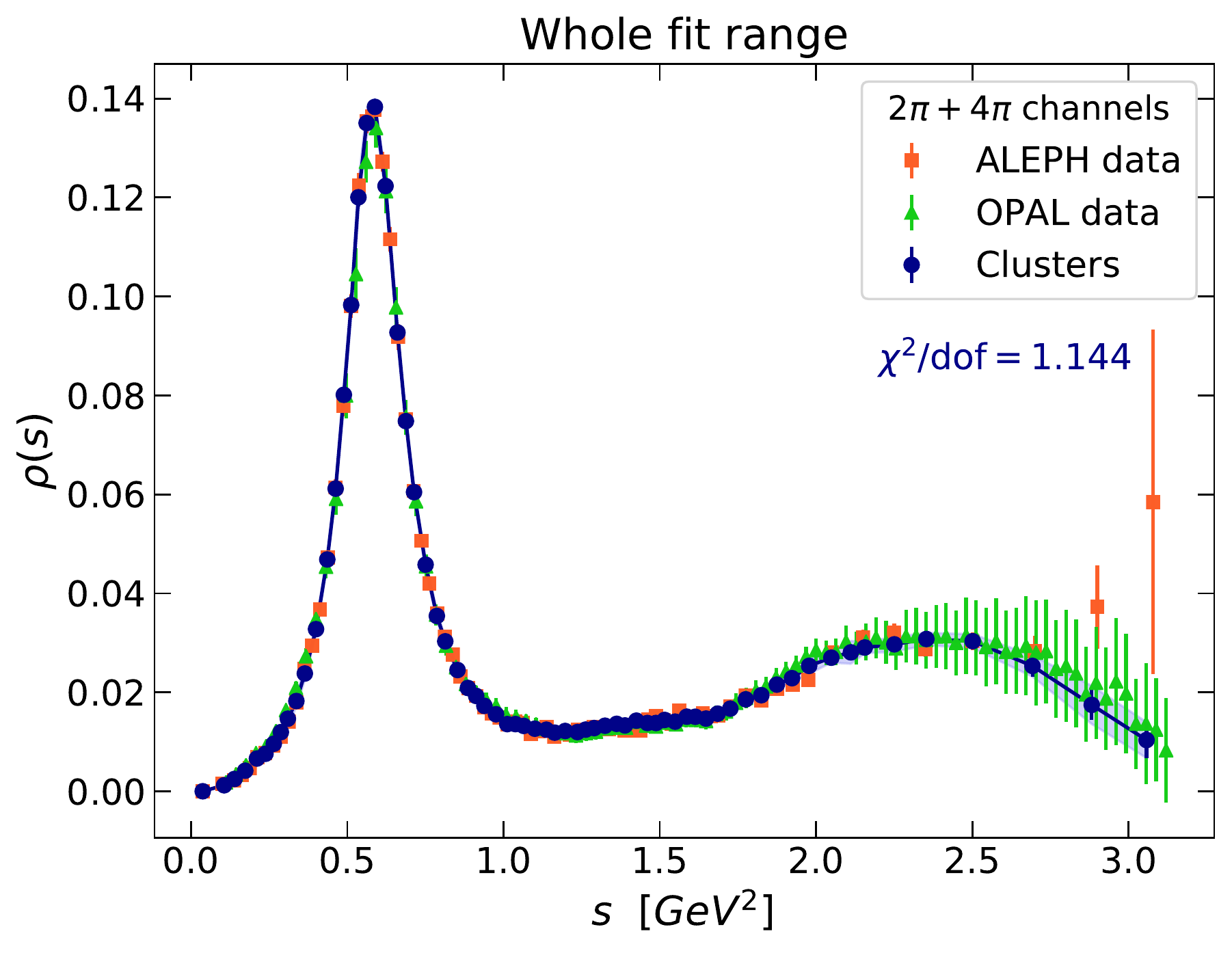}
\includegraphics[width=0.47\textwidth]{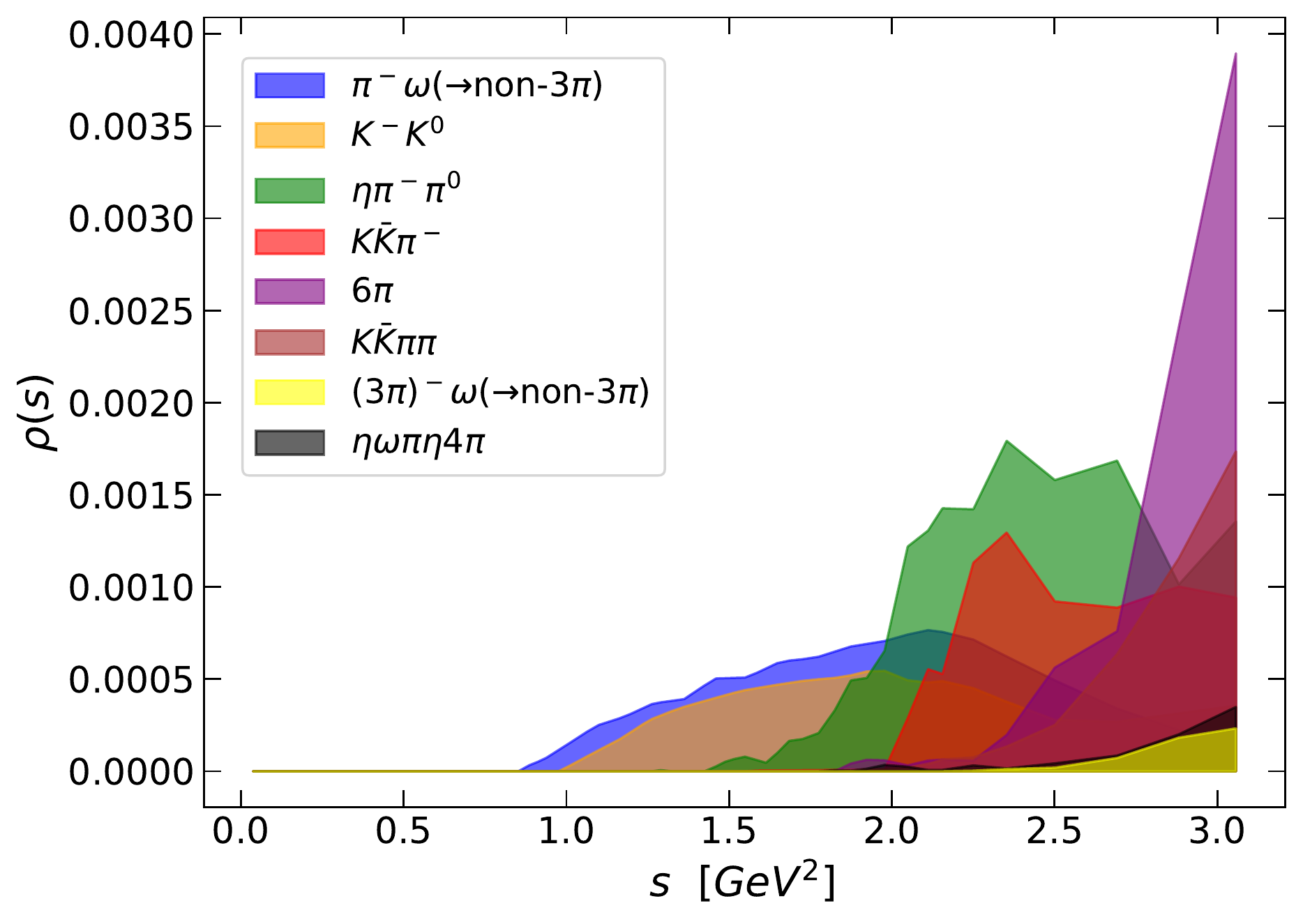}
\caption{(Left panel) ALEPH (orange), OPAL (green), and combined data (blue) for the sum of the dominant vector isovector channels: $\t\to\p^-\p^0\n_\t$, $\t\to 2\p^-\p^+\p^0\n_\t$, and $\t\to\p^-3\p^0\n_\t$. (Right panel) Residual mode contributions, channel by channel. }
\label{fig:DataComb}
\end{figure}

With the procedure outlined above one obtains the result displayed in the left panel of Fig.~\ref{fig:DataComb}. In this figure, we show the total ALEPH, OPAL, and combined data for $\t\to\p^-\p^0\n_\t$, $\t\to 2\p^-\p^+\p^0\n_\t$, and $\t\to\p^-3\p^0\n_\t$. These dominant channels are responsible for $98.0\%$ of the total inclusive V channel BF. The $\chi^2/dof$ of the combination, 1.144, is very reasonable and we have also checked that for each cluster, locally, the $\chi^2$ is, in general, very good. There are a handful of clusters with not-so-good $\chi^2$ but they are so few that local error inflation makes no difference in our final results.

The next step is the inclusion of the residual modes. We include, in our new spectral function, the following decay modes: (i) $\pi^- \omega (\rightarrow {\rm \mbox{non-}3}\pi)$, $K^- K^0$,
$\eta \pi^- \pi^0$, $K\bar{K}\pi$, $3\pi^- 2\pi^+ \pi^0$, and
$2\pi^- \pi^+ 3\pi^0$, which were also included by both the ALEPH and OPAL collaborations, (ii) $(3\pi )^- \omega(\rightarrow {\rm \mbox{non-}3}\pi)$
and $K\bar{K}\pi\pi$, included only by ALEPH, and (iii) the small additional contributions from $\pi^- 5\pi^0$ and
$\eta \omega\pi +\eta 4\pi$, not included in either of the ALEPH or OPAL analyses.
With the exception of the contributions from $\tau \to K^-K^0\nu_\tau$, which were measured by BaBar~\cite{BaBar:2018qry}, all the other modes are obtained from the respective $e^+e^-$ cross sections with CVC~\cite{BaBar:2007ceh,BaBar:2017zmc,Achasov:2016zvn,SND:2014rfi,Achasov:2017kqm,BaBar:2018erh,BaBar:2018rkc,Gribanov:2019qgw,BaBar:2006vzy,CMD-3:2013nph,Achasov:2019nws,CMD-3:2017tgb,Achasov:2019duv,BaBar:2007qju,Achasov:2016eyg}, applying the necessary corrections for vacuum polarization effects. The individual channels are normalized to the respective BFs described in the 2019 HFLAV compilation~\cite{HFLAV:2019otj} (correlations among these values are taken into account). Since the residual modes contribute at energies outside the region of narrow resonances, isospin-breaking corrections, which are expected to be of the order of a percent, are very small given the experimental errors and the smallness of the residual mode contributions to the inclusive spectral function. A much more detailed description of each of the individual residual modes can be found in Sec.~III.C of\cite{Boito:2020xli}. The contribution of the residual modes is shown, channel by channel, in the right-hand panel of Fig.~\ref{fig:DataComb}.

\begin{figure}[t]
\centering
\includegraphics[width=0.55\textwidth]{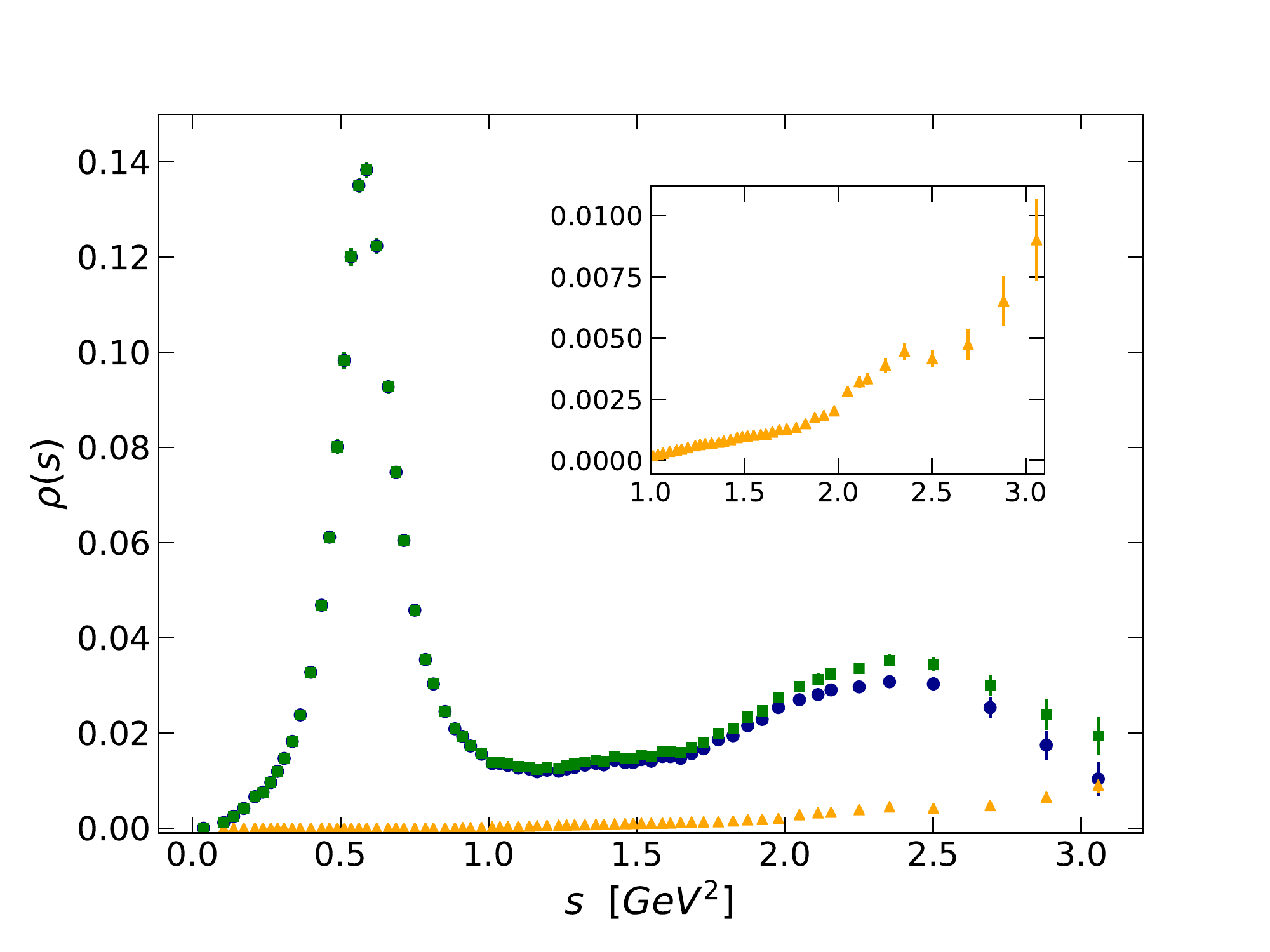}
\caption{Total vector isovector spectral function (green) obtained from the addition of the dominant modes (blue) and the residual modes (yellow).}
\label{fig:SpecFunc}
\end{figure}

The final result for the vector isovector spectral function is then obtained by the addition of the dominant and residual channels, and is shown in Fig.~\ref{fig:SpecFunc} (green data points). This new vector isovector spectral function, entirely based on experimental data, covers 99.95\% of the total inclusive vector isovector BF and has smaller uncertainties than its predecessors~\cite{Davier:2013sfa,OPAL:1998rrm}.

\section{Results for the strong coupling}

Here we describe the main results of the QCD finite-energy-sum-rule (FESR) analysis of Ref.~\cite{Boito:2020xli}. We follow the analysis strategy developed in~\cite{Boito:2011qt} and previously employed in the analysis of ALEPH and OPAL data~\cite{Boito:2012cr,Boito:2014sta}.
For further details of the analysis we refer to~\cite{Boito:2020xli} and to the previous publications~\cite{Boito:2011qt,Boito:2012cr,Boito:2014sta}.

As is well established~\cite{Braaten:1991qm}, the QCD analyses employ FESRs where, on the experimental side, one has integrated spectral function moments, obtained from a discretized integral over the data set, and on the theory side one has an integral of the relevant current-current correlation function, $\Pi(s)$, along a closed contour in the complex plane, such that the value of the coupling is kept in the perturbative
domain:
\begin{equation}
\frac{1}{s_0}\int_0^{s_0}ds\,w(s)\,\r(s)
=-\frac{1}{2\p i\, s_0}\oint_{|z|=s_0}
dz\,w(z)\,\P(z),
\end{equation}
where $\rho(s)$ is the total vector isovector spectral function of Fig.~\ref{fig:SpecFunc}.
In the FESRs any analytical weight function $w(s)$ can be employed and this freedom is used in order to emphasize or suppress the different theoretical contributions.
The QCD contributions on the theoretical side can be split into a perturbative component, cast in terms of the perturbative Adler function (known to $\alpha_s^4$~\cite{Baikov:2008jh,Herzog:2017dtz}), and non-perturbative contributions: the operator product expansion (OPE) condensates and duality violations (DVs), which go beyond the usual OPE. The results that we report here are based on Fixed Order Perturbation Theory (FOPT) (see~\cite{Beneke:2008ad}, for example) which can be more directly compared with determinations of $\alpha_s$ from other sources. We do not quote values for Contour Improved Perturbation Theory~\cite{LeDiberder:1992jjr} (CIPT) since there are solid indications of a potential inconsistency between this prescription and the standard OPE corrections~\cite{Hoang:2020mkw,Hoang:2021nlz}.

We follow the DV strategy, where one choses weight functions that strongly suppress higher-order OPE condensates, which avoids completely any problem related to the truncation of OPE series. (Assumptions about the truncation of the OPE can be dangerous in this context, and may lead to uncontrolled systematic effects~\cite{Boito:2016oam,Boito:2019iwh,Golterman}.) The price to pay is an unavoidable enhancement of DV contributions which must be included in the theoretical description. We use the parametrization of Refs.~\cite{Cata:2005zj,Boito:2017cnp} which is based on generally accepted assumptions about the QCD resonance spectrum. We then fit simultaneously several FESRs, in each case integrating the data from threshold up to $s_0$, with $s_0$ varied between $s_{\rm min}$ and its highest possible value (that of the right-most point in Fig.~\ref{fig:SpecFunc}, which is very close to $m_\tau^2$). The parameters of the fits are $\alpha_s$, the four DV parameters and, depending on the moment used, the relevant OPE condensates.

\begin{figure}[t]
\centering
\includegraphics[width=0.65\textwidth]{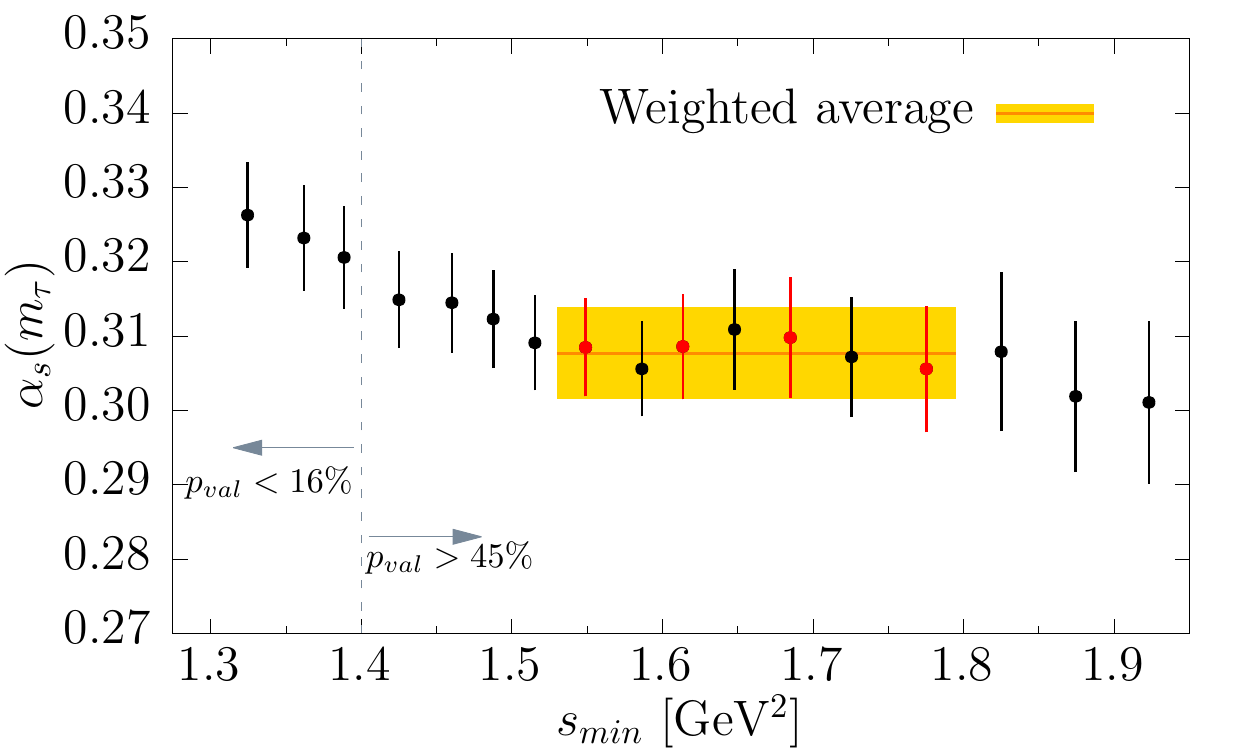}
\caption{$\alpha_s(m_\tau)$ as a function of $s_{\rm min}$ from the fits described in the text. The horizontal red line and the yellow band give the final result with uncertainties, Eq.(\ref{eq:alphasFOPTmtau}). Points in red are included in the final weighted average (see text).}
\label{fig:alphasRes}
\end{figure}

\begin{figure}[t]
\centering
\includegraphics[width=0.55\textwidth]{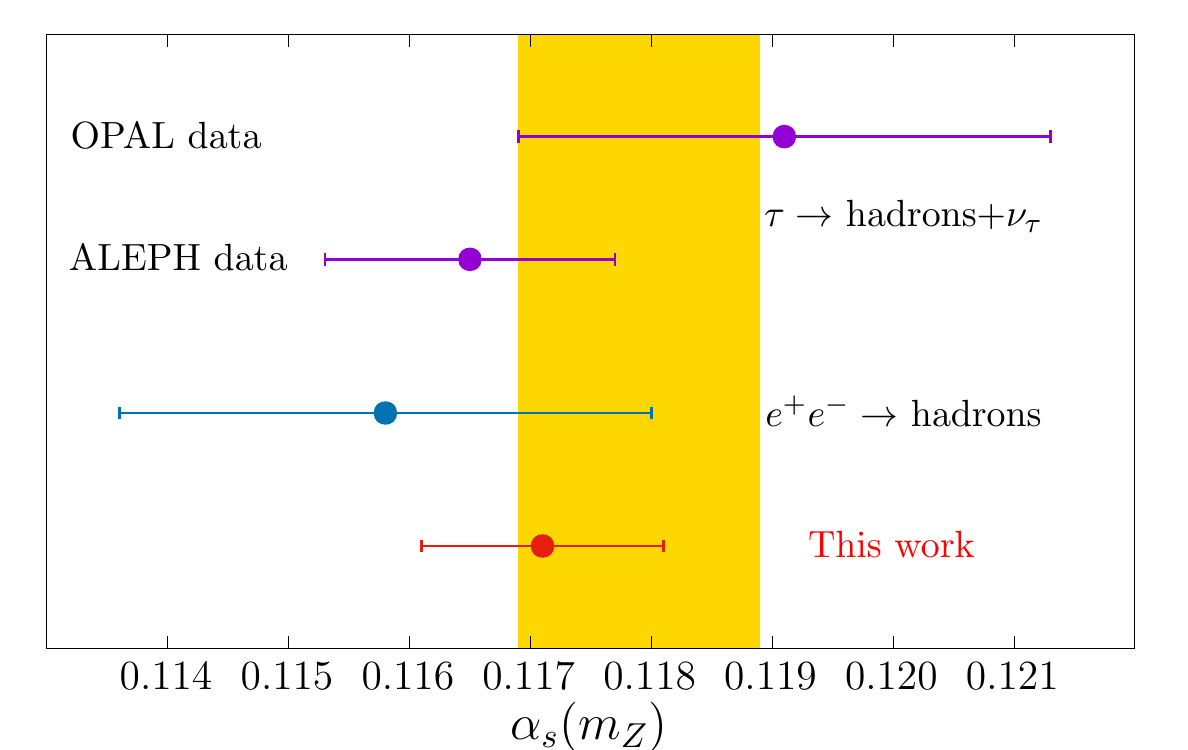}
\caption{Comparison of $\alpha_s(m_Z)$ within FOPT from the analyses of OPAL hadronic tau decay data~\cite{Boito:2012cr}, ALEPH hadronic tau decay data~\cite{Boito:2014sta}, from $e^+e^-\to {\rm hadrons}$ below charm threshold~\cite{Boito:2018yvl}, and from the new vector isovector spectral function~\cite{Boito:2020xli} (this work). The yellow band gives the Particle Data Group world average~\cite{ParticleDataGroup:2020ssz}.}
\label{fig:alphasComp}
\end{figure}

In Fig.~\ref{fig:alphasRes} we show results for $\alpha_s(m_\tau)$ from FESRs using $w(x)=1$ obtained from fits with different $s_{\rm min}$, as a function of $s_{\rm min}$. As one can see, as long as $s_{\rm min}$ is not too low, the results obtained are very consistent and stable. We have performed other consistency checks, such as using other weight functions $w(x)$, as well as using combinations of moments, always with very consistent results. To obtain a final value of $\alpha_s(m_\tau)$ we perform a weighted average of every second result within the yellow band of Fig.~\ref{fig:alphasRes}. In this average, the strong correlation between the different values of $\alpha_s(m_\tau)$ is properly taken into account. Our final result is then
\begin{eqnarray}
\a_s(m_\t)&=&0.3077\pm 0.0065_{\rm stat}\pm 0.0038_{\rm pert} \\
&=& 0.3077\pm 0.0075\qquad\qquad(\overline{{\rm MS}}, n_f=3\ ,\ \mbox{FOPT})\ \label{eq:alphasFOPTmtau}
\end{eqnarray}
where we quote a statistical (stat) and a perturbative (pert) error.\footnote{For details about the error associated with perturbation theory see Sec.~IV.C of~\cite{Boito:2020xli}.} Evolving this result to the $Z$-mass scale with five-loop running~\cite{Baikov:2016tgj} and four-loop matching~\cite{Schroder:2005hy,Chetyrkin:2005ia} we find
\begin{equation}
\alpha_s(m_Z)=0.1171\pm 0.0010 \qquad(\overline{{\rm MS}}, n_f=5). \label{eq:alphasFOPTmZ}
\end{equation}
This result is compared, in Fig.~\ref{fig:alphasComp}, with our previous determinations using the same strategy with other data sets~\cite{Boito:2012cr,Boito:2014sta,Boito:2018yvl}.

A comparison of our results with other analyses found in the literature is not completely straightforward, since they are based on different data sets and different analysis strategies. The results from Refs.~\cite{Davier:2013sfa,Pich:2016bdg} employ variants of the so-called ``truncated OPE strategy" (tOPE), aimed at suppressing the DV contributions, but in which several OPE condensates have to be neglected in the fits. This is a strong assumption and there is evidence that it leads to an uncontrolled systematics~\cite{Boito:2016oam,Boito:2019iwh,Golterman}. With this caveat in mind, the results of~\cite{Davier:2013sfa,Pich:2016bdg} are significantly higher than ours (and than the PDG world average~\cite{ParticleDataGroup:2020ssz}) and read $\alpha_s(m_Z)=0.1199\pm 0.0015$~\cite{Davier:2013sfa} and $\alpha_s(m_Z)=0.1197\pm 0.0015$.~\cite{Pich:2016bdg}. A more recent analysis~\cite{Ayala:2021mwc}, which employs a variant of the tOPE strategy that has not yet been subjected to the same level of scrutiny as those of Refs.~\cite{Boito:2012cr,Boito:2014sta,Davier:2013sfa,Pich:2016bdg}, gives a value of $\alpha_s$ from FOPT essentially identical to the one we quote in Eq.~(\ref{eq:alphasFOPTmZ}).

\section{Conclusion}
We have recently produced a new vector isovector spectral function based on a combination of the ALEPH and OPAL spectra for the dominant $\pi\pi^0$, $\pi 3\pi^0$, and $3\pi \pi^0$ $\tau$ decay channels, supplemented by the addition of contributions from all numerically significant residual decay modes. The residual modes contributions are obtained entirely from experimental data, using the recent BaBAR spectrum for $\tau \to K^-K^0\nu_\tau$~\cite{BaBar:2018qry} and a wealth of recent CVC-related cross-section results from $e^+e^-\to {\rm hadrons}$ obtained by CMD-3, SND and BaBar~\cite{BaBar:2007ceh,BaBar:2017zmc,Achasov:2016zvn,SND:2014rfi,Achasov:2017kqm,BaBar:2018erh,BaBar:2018rkc,Gribanov:2019qgw,BaBar:2006vzy,CMD-3:2013nph,Achasov:2019nws,CMD-3:2017tgb,Achasov:2019duv,BaBar:2007qju,Achasov:2016eyg}. This new spectral function is more inclusive than its predecessors, since it contains more residual modes, is solely based on experimental data (with no Monte Carlo simulated channels), and has smaller errors. Isospin-breaking corrections for the CVC related channels are expected to be very small given the size of experimental errors and the smallness of the contributions from the residual modes. It is important to observe that an analogously improved spectral function cannot be built for the axial channel, since the $e^+e^-\to {\rm hadrons}$ cross-sections are purely vector at this energy scale.

We analyze FESRs based on this new data set in order to extract the value of the strong coupling at the tau mass scale, $\alpha_s(m_\tau)$. Our result, $\alpha_s(m_\tau)=0.3077\pm 0.0075$, is very competitive and corresponds to $\alpha_s(m_Z)=0.1171\pm 0.0010$. The results of the present analysis could be improved with new measurements for the spectral distributions of the dominant vector $\tau$ decay channels.

\section*{Acknowledgements}
We would like to thank the organizers of the Tau2021 workshop for keeping alive this conference series during difficult times.
We thank Alex Keshavarzi for valuable discussions.

\paragraph{Funding information}
DB was supported by the S\~ao Paulo Research Foundation (FAPESP)
Grant No.~2015/20689-9, by CNPq Grant No.~309847/2018-4, and by Coordena\c c\~ao de Aperfei\c coamento de Pessoal de N\'ivel Superior – Brasil (CAPES) – Finance Code 001.
MG and WS are supported by the U.S.\ Department of Energy,
Office of Science, Office of High Energy Physics, under Award No.
DE-SC0013682.
MVR is supported by FAPESP grant No.~2019/16957-9.
KM is supported by a grant from the Natural Sciences and Engineering
Research Council of Canada.
SP is supported by CICYTFEDER-FPA2017-86989-P and by Grant No. 2017 SGR 1069.
IFAE is partially funded by the CERCA program of the Generalitat de Catalunya.

\bibliography{References}

\nolinenumbers

\end{document}